\documentclass[10pt,twocolumn,letterpaper]{article}

\usepackage{iccv}              

\usepackage[accsupp]{axessibility}  

\definecolor{iccvblue}{rgb}{0.21,0.49,0.74}
\usepackage[pagebackref,breaklinks,colorlinks,allcolors=iccvblue]{hyperref}

\def\paperID{6} 
\def\confName{ICCV}
\def\confYear{2025}

\title{Controllable Single-shot Animation Blending with Temporal Conditioning}

\author{Eleni Tselepi\\
ECE Dept., Univ. of Thessaly,\\
Volos, Greece \\
{\tt\small etselepi@uth.gr}
\and
Spyridon Thermos\\
Moverse,\\
Thessaloniki, Greece \\
{\tt\small spiros@moverse.ai}
\and
Gerasimos Potamianos\\
ECE Dept., Univ. of Thessaly,\\
Volos, Greece \\
{\tt\small gpotamianos@uth.gr}
}

\begin{document}
\twocolumn[{%
\renewcommand\twocolumn[1][]{#1}%
\maketitle

\begin{center}
    \centering
    \includegraphics[width=\textwidth]{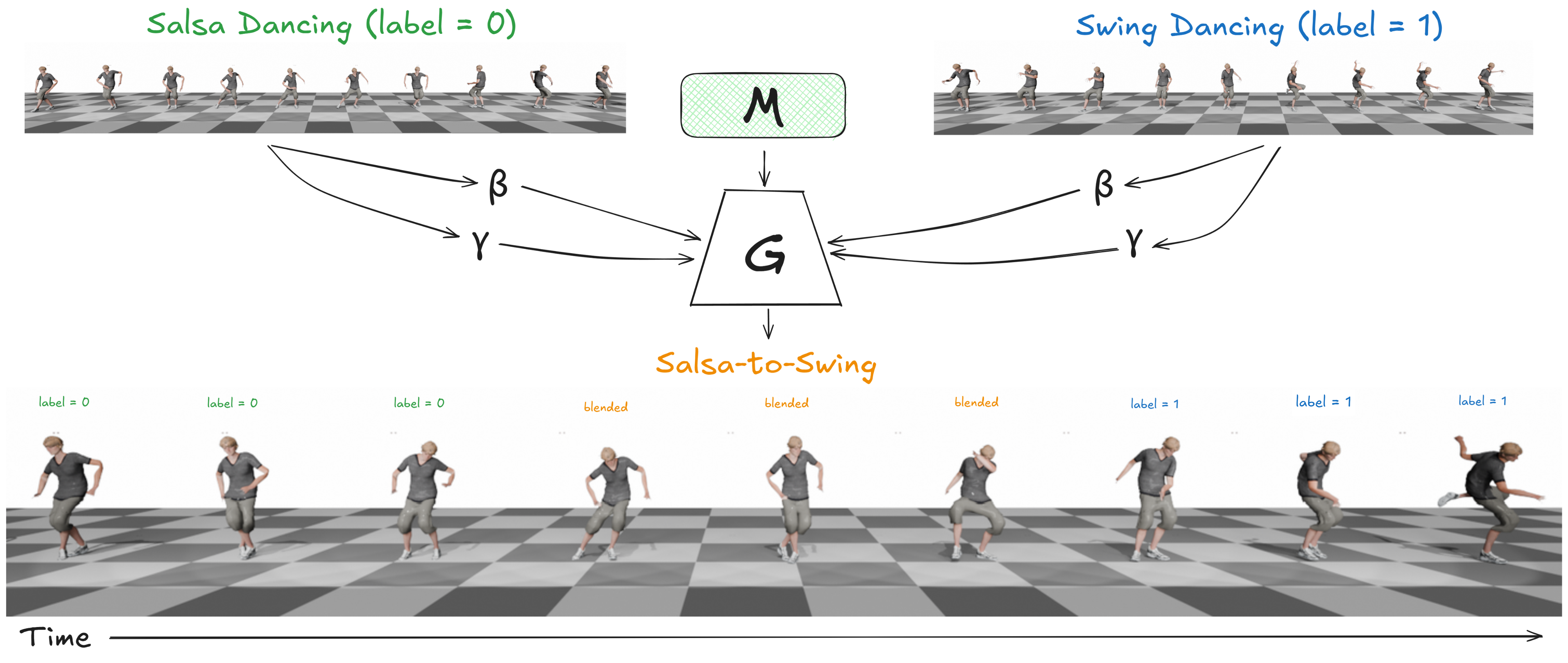}
    \captionof{figure}{
        We introduce a single-shot motion blending approach that uses a batched version of the GANimator~\cite{li2022ganimator}, extended with a SPADE-like \cite{spade} conditioning mechanism. Our method enables on-the-fly blending of two or more input animations by scaling ($\gamma$) and shifting ($\beta$) the motion features ($\mathcal{M}$) of the skeleton convolutions~\cite{skeleton_conv}, generating ($G$) coherent and temporally consistent motion transitions in a single forward pass, without requiring retraining or large motion datasets. Best viewed in color.
    }
    \label{fig:teaser}
\end{center}
}]
\begin{abstract}
Training a generative model on a single human skeletal motion sequence without being bound to a specific kinematic tree has drawn significant attention from the animation community. Unlike text-to-motion generation, single-shot models allow animators to controllably generate variations of existing motion patterns without requiring additional data or extensive retraining. However, existing single-shot methods do not explicitly offer a controllable framework for blending two or more motions within a single generative pass.
In this paper, we present the first single-shot motion blending framework that enables seamless blending by temporally conditioning the generation process. Our method introduces a skeleton-aware normalization mechanism to guide the transition between motions, allowing smooth, data-driven control over when and how motions blend.
We perform extensive quantitative and qualitative evaluations across various animation styles and different kinematic skeletons, demonstrating that our approach produces plausible, smooth, and controllable motion blends in a unified and efficient manner. Our project page can be found \href{https://elenitselepi.github.io/BlendAnim-project/}{here.}
\end{abstract}    
\section{Introduction}
\label{sec:intro}
Creating expressive and believable character animation has long been critical to visual storytelling in art, film, and interactive media. A key enabler of this is animation blending, \textit{i.e.,} the process of smoothly transitioning between distinct motion segments to produce continuous movement. Early research \cite{aksan02, aksan03, kovar02} laid the foundations for blending as a form of creative motion composition, allowing artists to navigate, recombine, and interpolate motion sequences in a controllable way. These early data-driven techniques helped creators prototype complex behaviors and stylized movements without authoring every frame manually.

The advent of deep generative models has expanded the possibilities of generating motion, particularly through text-to-motion paradigms that map natural language to dynamic character movements~\cite{ghosh2021,petrovic2022,karunratanakul2023}. Although these models offer expressive capabilities and high realism, they also introduce significant barriers to end-users such as artists, as they require large-scale annotated datasets, high computational cost, and familiarity with generative AI and prompt engineering.
In this context, single-shot animation generation offers a more accessible and efficient alternative. By synthesizing entire animation sequences in a single forward pass, these models bypass the need for extensive training corpora and enable IP-free generation from individual motion samples. In particular, when implemented with lightweight architectures such as Generative Adversarial Networks (GANs)~\cite{gans2014}, these approaches support fast, local inference, making them ideal for deployment on modest hardware and enabling integration into real-time or resource-constrained creative environments. 

However, current single-shot approaches lack the ability to blend multiple motion styles or sequences in a controllable and interpretable way. In this work, we address this gap by introducing a single-shot motion blending model that enables seamless blending by temporally conditioning the generation process. Our method incorporates a SPADE-inspired conditioning \cite{spade} scheme tailored for skeleton-aware convolutions \cite{skeleton_conv}, enabling 3D artists and animators to blend multiple input motions into coherent and temporally consistent animations in a single generative pass (see also Fig.~\ref{fig:teaser}). This opens new possibilities for art-directed motion synthesis, where creators can combine and remix motion sources to create the desired character performances without needing complex pipelines.

To our knowledge, this is the first approach that enables controllable animation blending using a single-shot model. Our contributions are the following: 
\begin{itemize}
    \item we propose a conditional blending framework that leverages the inherent batching in GANimator to support multi-source input control,
    \item we demonstrate that our approach enables expressive, interpretable, and high-fidelity blending of motions, without the need for large datasets or complex post-processing.
\end{itemize}

The rest of the paper is organized as follows: Prior works in conditional motion generation and motion blending are discussed in Sec.~\ref{sec:related_work}; a detailed background of the single-shot backbone architecture along with our blending approach are covered in Sec.~\ref{sec:method}; the experimental results are presented in Sec.~\ref{sec:experiments}; and finally our conclusions are reported in Sec.~\ref{sec:conclusion}.

\section{Prior Art}
\label{sec:related_work}
\textbf{Conditional motion generation}. From early rule-based and physics-driven systems to modern deep generative models, the process of conditioning the human body motion generation has evolved significantly. The foundational work of Badler \textit{et al.}~\cite{simulating93} introduces structured frameworks for animation and control, using kinematic constraints and procedural logic to simulate believable human behaviors, laying the groundwork for later data-driven approaches. With the advent of deep learning, models like the Neural State Machine~\cite{starke2019neural} began conditioning motion generation based on scene and interaction context, using discrete state transitions learned from data to guide character behavior in dynamic environments. Building on this, Ling \textit{et al.}~\cite{ling2020character} introduced control over motion generation through latent space manipulation, allowing fine-grained pose or trajectory conditioning within a probabilistic generative framework. More recently, Athanasiou \textit{et al.}~\cite{athanasiou2022teach} presented TEACH, a transformer-based architecture that composes human motion directly from textual descriptions, marking a shift toward high-level semantic control of motion synthesis. The aforementioned works illustrate the progression from explicit control schemes to conditioned generative models, moving toward more intuitive and expressive interfaces for animating human motion.

\textbf{Motion blending.}
Motion blending has been a fundamental technique in character animation, primarily implemented as temporal composition to ensure smooth and realistic transitions between distinct motions. This approach emphasizes not only the technical stitching of sequences \cite{aksan02, kovar02} but also the perceptual realism of transitions \cite{kovar04,mukai05}, which is critical for maintaining naturalness in movement. More recently, Athanasiou~\etal~\cite{athanasiou2022teach} investigate blending methods that prioritize temporal continuity to preserve motion plausibility during transitions, while Barquero~\etal.~\cite{seamless2024} introduce techniques that adaptively align motion fragments to enhance transition quality, reinforcing the importance of timing and kinematic coherence. Shafir~\etal~\cite{shafir2024human} further investigate this direction by proposing transition strategies that focus on perceptual smoothness, enabling smoother and believable motion sequences. Such works underscore that motion blending has largely been addressed as a problem of temporal composition, with a strong emphasis on the realism and quality of the transition.

\textbf{In-betweening.}
Animation in-betweening, a variant of motion blending, focuses on generating intermediate frames that connect key poses to create smooth and coherent motion sequences. Recent works have approached this task using deep learning techniques, often relying on multi-stage or sequential architectures. Both \cite{harvey20} and \cite{qin22} employ hierarchical or cascaded models to incrementally generate intermediate motions while preserving temporal consistency and motion semantics. Oreshkin~\etal~\cite{oreshkin24} propose a method to model the motion changes between frames, enabling more precise interpolation of human movement. Such methods generally rely on multiple inference steps or recurrent structures to maintain continuity. In contrast, Duan~\etal~\cite{duan21} propose a single-pass transformer-based framework to generate the entire in-between sequence, offering faster and more efficient synthesis. This single-shot strategy is conceptually aligned with our proposed GAN-based single-shot blending approach, which similarly aims to generate realistic transitions in a single inference step, bridging the gap between motion in-betweening and real-time blending applications.

\textbf{Single-shot models.}
Single-shot generation has emerged as a compelling approach for generating diverse outputs from a single example, initially in the image domain and more recently extending to more complex data representations like 3D scenes and motion data. Early works like InGAN~\cite{shocher2019ingan} demonstrated the feasibility of conditional generative models trained on a single image for tasks such as remapping. SinGAN~\cite{rottshaham2019singan} introduced unconditional single-image generation through a multi-level progressive learning scheme, employing a patch-based discriminator~\cite{li2016patch,isola2017cvpr} and a reconstruction loss to stabilize training and prevent mode collapse. These concepts were extended to 3D content with Sin3DM~\cite{wu2023sin3dm}, which utilized diffusion models to overcome runtime constraints, and SinGRAF~\cite{son2023singraf} that applied neural rendering to generate 3D scene variations from a single sample. In the domain of motion, methods like GANimator~\cite{li2022ganimator} and SinMDM~\cite{SinMDM} adapted adversarial and diffusion-based strategies for generating motion from a single sequence. Of particular relevance is BlendGAN~\cite{blendgan}, which introduces a framework for learning and blending internal distributions of single images through spatially conditioned GANs. Inspired by the aforementioned work, we propose a method for transferring the single-sample generation concept to the controllable human motion blending domain. Specifically, we propose a single-shot motion blending approach that learns to generate smooth and realistic transitions between two or more motions at the desired time-frame by modeling and blending their internal distributions, without the need for large datasets or extreme computational resources.
\section{Method}
\label{sec:method}


\begin{figure}
    \centering
    \includegraphics[width=1\columnwidth]{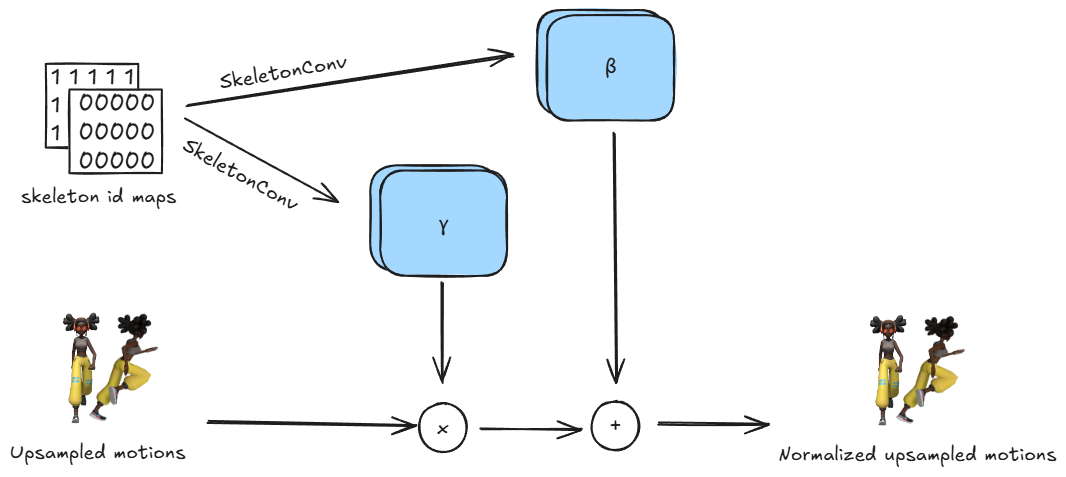}
    \caption{A SPADE-like block that receives the temporal conditioning information and projects it onto an embedding space using two skeleton-aware convolutional layers to produce the modulation tensors $\gamma$ and $\beta$, which are used to normalize the motion representation.}
    \label{fig:spade}
\end{figure}

\subsection{Background}
\label{subsec:backgound}
To build upon the GANimator foundation laid by Li~\etal~\cite{li2022ganimator}, we first revisit its motion data representation and formalize the relevant notations. We represent a motion sequence as a matrix $\mathbf{M} \in \mathbb{R}^{T \times D}$, where $T$ is the number of frames and $D = JQ+C+3$ is the feature dimension. Here, $J$ is the number of skeletal joints, $Q=6$ is the 6D rotation parameterization for joint orientations, $C$ represents foot contact features, and the final 3 dimensions correspond to the root joint's vertical position and planar velocity. For the adopted temporal hierarchical multi-stage scheme, we denote the downsampled motion representation at stage $i \in \{1, \dots, S\}$ as $\mathbf{M}^{(i)} \in \mathbb{R}^{T_i \times D}$, where $T_i$ is the number of frames at that resolution.

\textbf{GANimator}. This model adopts a coarse-to-fine temporal hierarchical scheme for generating motion, where learning is structured across $S=7$ stages, each consisting of a generator-discriminator $(G, D)$ pair, split among four temporal resolution levels. Rather than training all pairs simultaneously in an end-to-end fashion, the model employs a progressive training strategy. At each of the four levels, two stage pairs ${G(\cdot), D(\cdot)}$ are responsible for learning to generate motion representations at the specific temporal resolution of that level. Lower levels generate coarse motion patterns that are later refined by higher levels, while the final level contains only one ${G(\cdot), D(\cdot)}$ pair and is responsible for generating the final motion features at the desired resolution. This design allows the model to first learn global, low-resolution motion structures, and then progressively refine them, capturing finer motion dynamics as it moves up the temporal resolution pyramid. Such a strategy improves training stability and enables realistic motion synthesis from a single motion sequence, without the need for large datasets. However, the original implementation of GANimator does not support mini-batch training, which is necessary in single-shot blending, as the model learns to assign different semantic labels to the different motions represented in each batch. For this reason we employ the GANimator variant proposed in \cite{roditakis2024singleshot}, which allows us to train the single-shot GAN using a batch size equal to the number of motions we want to blend.

\subsection{Single-shot blending}

\begin{figure*}
    \centering
    \includegraphics[width=\textwidth]{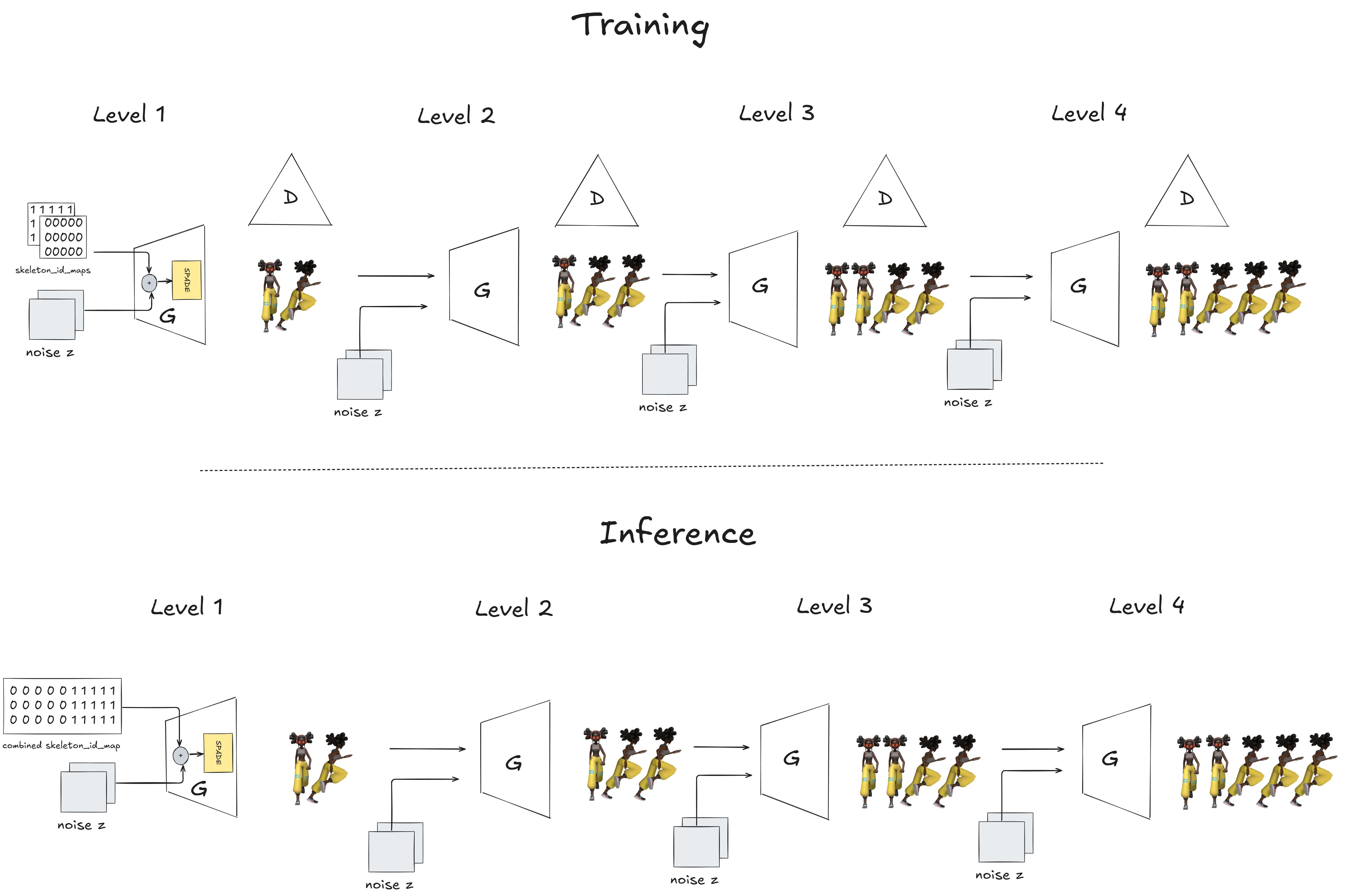}
    \caption{A schematic representation of the training (top) and inference (bottom) processes. During training each semantic label is assigned to a different batch of the input motions, while at inference a single sequence of labels drives the generation (blending) process. For simplicity, a single stage of $(G, D)$ pairs (training) or generators (inference) is shown at each of the four levels.}
    \label{fig:training-testing}
\end{figure*}

To enable conditional blending, the generators of the first level (\textit{i.e.,} the ones of the first two ${G(\cdot), D(\cdot)}$ pairs) incorporate a Spatially-Adaptive Denormalisation (SPADE) layer \cite{spade} with a skeleton-aware convolutional structure \cite{skeleton_conv}. In particular, these generators $G(\cdot)$ receive a white Gaussian noise map and a skeleton identity tensor that has the same number of frames as the level being processed and the same number of channels as the motion sequences. Each channel contains a one-hot representation identifying the specific motion sequence. The SPADE layers of the generators use this tensor to compute position-dependent modulation parameters $\gamma$ and $\beta$ for every spatial location. As depicted in Fig.~\ref{fig:spade}, the SPADE block of the generators receives a skeleton identity tensor \textit{skeleton\_id\_map} and projects it onto an embedding space using two skeleton-aware convolutional layers to produce the modulation parameters (tensors) $\gamma$ and $\beta$. These parameters are then used to scale ($\gamma$) and shift ($\beta$) the motion representation of previous generators.

Unlike the traditional SPADE design, which applies normalization to the input using batch or instance normalization before performing scale and shift operations, we omit normalization in our model. Our generator is designed to blend a number of motion frames rather than synthesizing visual textures. Removing normalization, which typically captures high-frequency visual signals such as texture or style, helps preserve the fundamental structure of the input motion, which is critical for smooth transitions. The hierarchical architecture of the GAN helps to learn to distinguish between the two input motions at coarser levels, leading to more effective blending in the final output.

\subsubsection{Loss functions}
The training loss consists of three components: the adversarial, the reconstruction, and the foot contact consistency loss.

The adversarial loss for training the generators and discriminators is computed using the Wasserstein variant from \cite{wasserstein_gan}:

\begin{multline}
\mathcal{L}_{\text{adv}} = \min_{G_i} \max_{D_i} \mathbb{E} \left[ D_i(\mathbf{M}^{(i)}) - D_i(\hat{\mathbf{M}}^{(i)}) \right] \\
+ \lambda_{\text{gp}} \mathbb{E} \left[ \left( \left\| \nabla D_i(\tilde{\mathbf{M}}^{(i)}) \right\|_2 - 1 \right)^2 \right],
\label{eq:gan_loss}
\end{multline}

\noindent where $i \in [1,\dots, S]$ denotes the model's stage and $ \tilde{\mathbf{M}}^{(i)} = \alpha \hat{\mathbf{M}}^{(i)} + (1 - \alpha) \mathbf{M}^{(i)}$ is a linear combination of the generated motion $\hat{\mathbf{M}}^{(i)}$ and the ground truth $\mathbf{M}^{(i)}$. The second term, the gradient penalty, offers more stable training.

The reconstruction loss ensures that the model does not collapse to generating only a limited subset of movements and can reconstruct real motion sequences from a fixed latent vector, so the generators are supervised by an L1 loss:
\begin{equation}
\mathcal{L}_{\text{rec}} = \left\| G_i(\hat{\mathbf{M}}^{(i-1)}, z_i^*) - \mathbf{M}^{(i)} \right\|_1,
\end{equation}
\noindent where $z_i^*$ is a predefined noise embedding used to learn the reconstruction of $\mathbf{M}^{(i)}$.

The foot contact consistency loss ensures realistic foot contact in the generated motions:
\begin{equation}
\mathcal{L}_{\text{con}} = \frac{1}{T |\mathcal{F}|} \sum_{j \in \mathcal{F}} \sum_{t=1}^{T} \left\| \mathbf{v}_{t, j} \right\|_2^2 \cdot \sigma(C_{t,j}),
\end{equation}

\noindent where $\mathbf{v}_{t,j}$ is the velocity of foot joint $j$ at time $t$, $\mathcal{F}$ is the set of foot joints, $C$ denotes the contact features of the foot joints, and $\sigma(\cdot)$ represents a skewed Sigmoid function to estimate the contact probability.

Our overall loss for the final training stage ($i=S$) is summarized as:
\begin{equation}
\mathcal{L} = \lambda_{\text{adv}} \mathcal{L}_{\text{adv}} + \lambda_{\text{rec}} \mathcal{L}_{\text{rec}} + \lambda_{\text{con}} \mathcal{L}_{\text{con}},
\label{eq:total_loss}
\end{equation}

\noindent where $\lambda_{\text{adv}}$, $\lambda_{\text{rec}}$, and $\lambda_{\text{con}}$ are weighting coefficients that balance the contribution of each loss component. For the intermediate stages ($i<S$), the training objective omits the foot contact term ($\mathcal{L}_{\text{con}}$).

\subsubsection{Training and Inference}

Our model is trained once over all input motions, with each motion sequence provided in a separate batch. This is in contrast to prior single-shot GAN models such as GAN\-imator \cite{li2022ganimator}, which require training a distinct model for each individual motion. Attempts to train one of these models on several input motions simultaneously result in poor results, as can be seen in our ablation experiments (Table~\ref{tab:quantitative_results} and Table~\ref{tab:100_style_quantitative_results}). Training is performed stage-by-stage, where each generator and discriminator pair is trained independently while keeping the other stages fixed. During training, our model receives a \textit{skeleton\_id\_map} where each motion is assigned a unique identifier. This map allows the model to associate specific motion characteristics with corresponding identity channels.

At inference time, the \textit{skeleton\_id\_map} can be dynamically modified to control the generated output. We use a combined map that encodes both input motions, allowing the model to blend or switch between motions depending on the configuration of the \textit{skeleton\_id\_map}. As the map changes, the generator adapts accordingly, enabling conditional blending.

The training and inference processes are schematically represented in Fig.~\ref{fig:training-testing}, where the first motion is assigned a skeleton identity map filled with zeros, and the second with ones. During inference, the \textit{skeleton\_id\_map} is combined and assigned the first half of the frames to zero and the second half to one, resulting in a smooth transition that begins with the first motion and transitions into the second.
\section{Experiments}
\label{sec:experiments}

\subsection{Experimental setup}

\begin{figure}
    \centering
    \begin{minipage}{1\columnwidth}
        \centering
        \includegraphics[width=\columnwidth]{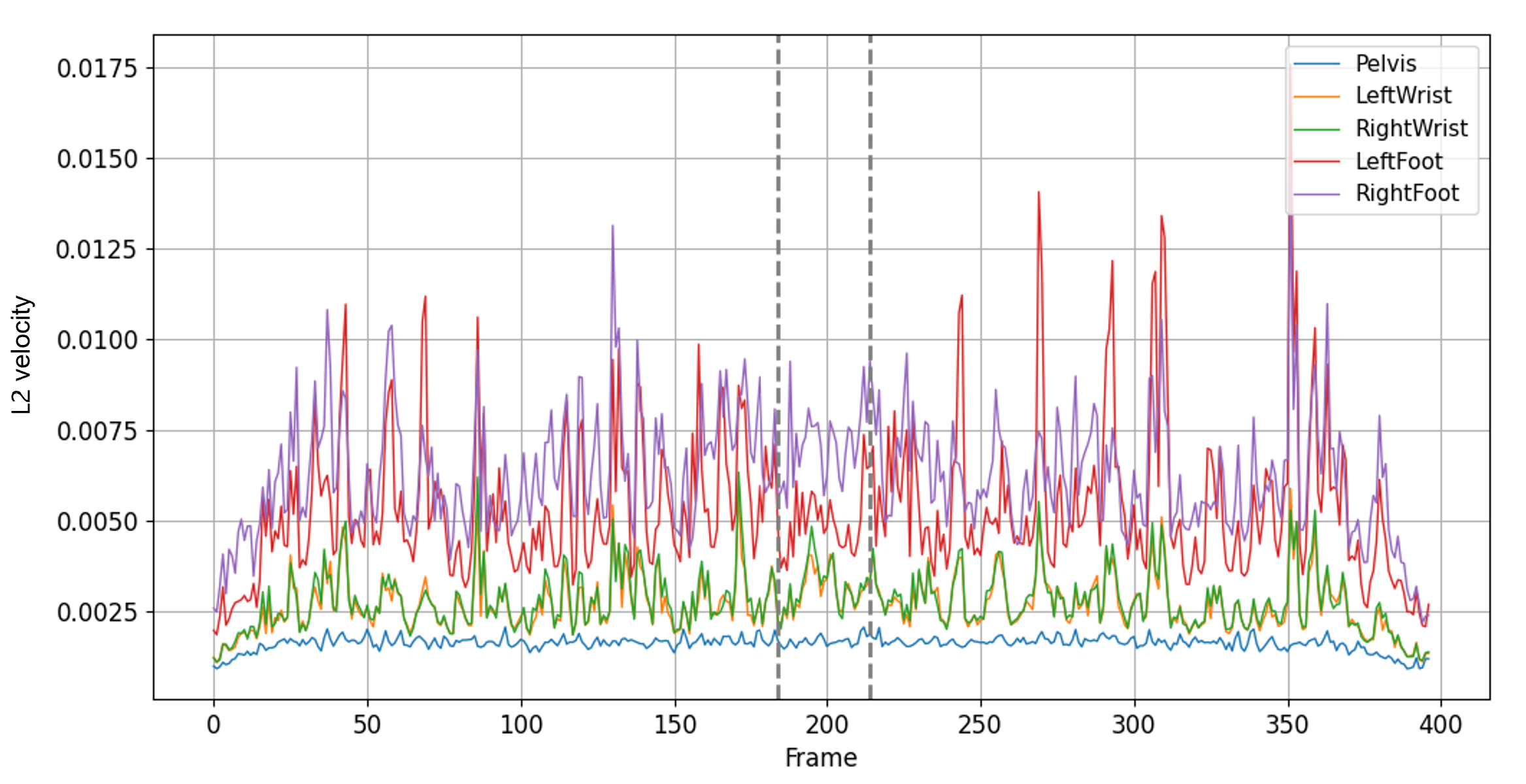}
    \end{minipage}
    \hspace{1cm}
    \begin{minipage}{1\columnwidth}
        \centering
        \includegraphics[width=\columnwidth]{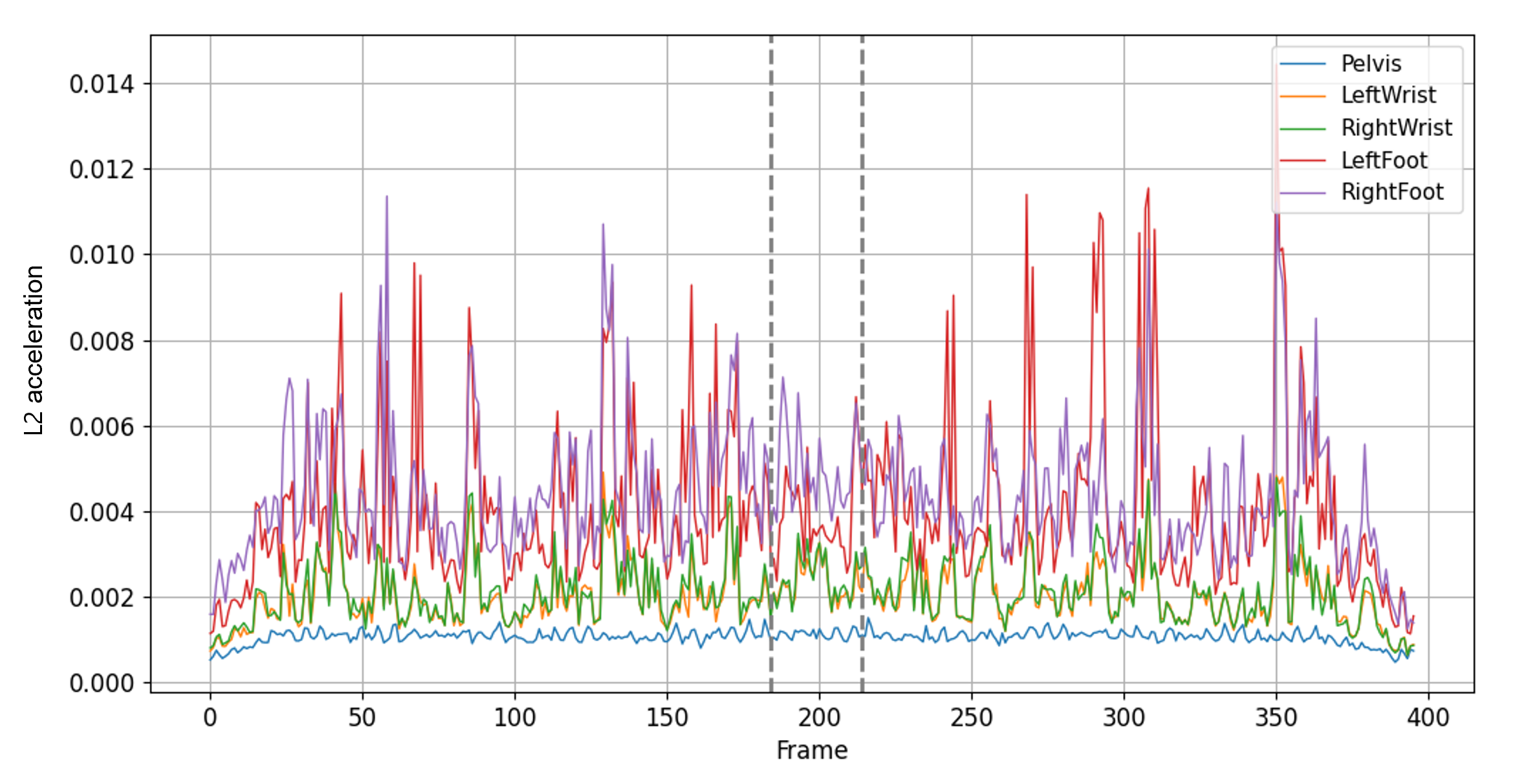}
    \end{minipage}
    \caption{L2 velocity (top) and acceleration (bottom) per joint for a blended motion from the 100STYLE dataset. The dotted vertical lines indicate the transition window where the two motions are merged. The relatively low values within this region suggest a smooth transition between the two input motions.}
    \label{fig:l2_metrics}
\end{figure}
\begin{table*}[t]
\centering
\caption{Ablation study evaluating different modulation mechanisms (SPADE and FiLM) and their integration at various levels of the generation hierarchy. We also provide a quantitative comparison with GANimator~\cite{li2022ganimator}. We report results on the Mixamo benchmark, blending ``Salsa\_Dancing'' and ``Swing\_Dancing'' motions.}
\label{tab:quantitative_results}
\begin{tabular}{lcccccc}
\toprule
& \textbf{FID (↓)} & \textbf{Cov. (↑)} & \textbf{GDiv. (↑)} & \textbf{LDiv. (↑)}
& \textbf{Inter-Div. (↑)} & \textbf{Intra-Div. (↓)}\\
\midrule
2\textsuperscript{nd} Level                        & 0.21 & 0.62 & 1.91 & 1.78 & 0.65 & 0.30 \\
3\textsuperscript{rd} Level                        & 0.23 & 0.59 & 1.92 & \textbf{1.80} & 0.65 & 0.32 \\
4\textsuperscript{th} Level                        & 0.24 & 0.57 & \textbf{1.93} & \textbf{1.80} & \textbf{0.68} & \textbf{0.29} \\
FiLM ~\cite{FiLM}                & 0.26 & 0.95 & 1.65 & 1.53 & 0. 60& 0.31 \\
GANimator~\cite{li2022ganimator} & 0.21 & 0.56 & \textbf{1.93} & \textbf{1.80} & \textbf{0.68} & \textbf{0.29} \\
Ours                             & \textbf{0.19} & \textbf{0.97} & 1.56 & 1.44 & 0.60 & 0.38 \\
\bottomrule
\end{tabular}
\end{table*}
\begin{table*}[t]
\centering
\caption{Ablation study results evaluated on the 100\_STYLE dataset, blending ``Chicken\_FW'' and ``Robot\_FW'' motions.}
\label{tab:100_style_quantitative_results}
\begin{tabular}{lcccccc}
\toprule
& \textbf{FID (↓)} & \textbf{Cov. (↑)} & \textbf{GDiv. (↑)} & \textbf{LDiv. (↑)}
& \textbf{Inter-Div. (↑)} & \textbf{Intra-Div. (↓)}\\
\midrule
2\textsuperscript{nd} Level                        & \textbf{0.09} & 0.50 & \textbf{2.06} & 1.98 & 0.48 & 0.37 \\
3\textsuperscript{rd} Level                        & \textbf{0.09} & 0.50 & \textbf{2.06} & \textbf{1.99} & \textbf{0.50} & 0.39 \\
4\textsuperscript{th} Level                        & 0.10 & 0.51 & 2.04 & 1.97 & \textbf{0.50} & 0.34 \\
FiLM ~\cite{FiLM}                & 0.14 & \textbf{1.00} & 0.50 & 0.44 & 0.33 & \textbf{0.29} \\
GANimator~\cite{li2022ganimator} & \textbf{0.09} & 0.51 & 2.04 & 1.96 & \textbf{0.50} & 0.34 \\
Ours                             & 0.13 & \textbf{1.00} & 0.51 & 0.46 & 0.35 & 0.34 \\
\bottomrule
\end{tabular}
\end{table*}

\begin{figure*}
    \centering
    \includegraphics[width=0.8\textwidth]{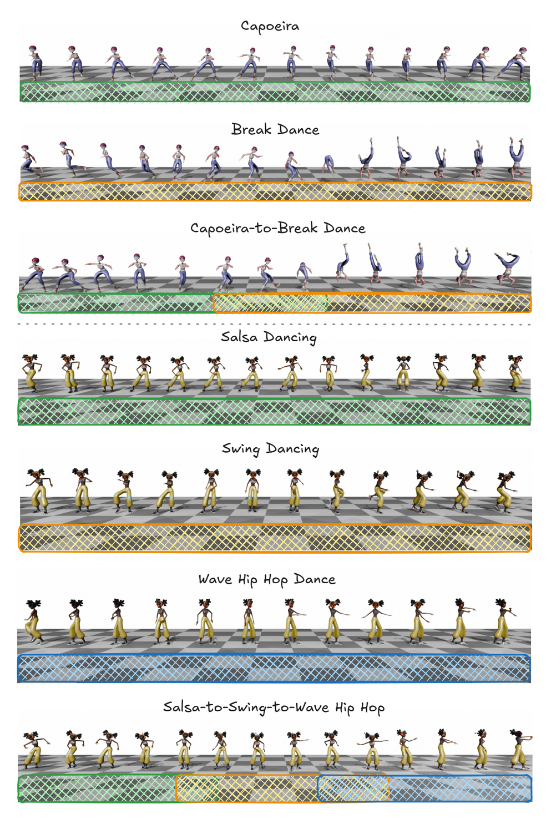}
    \captionof{figure}{
        The result of blending two (top) and three (bottom) animations from the Mixamo set \cite{Mixamo} with the proposed method. The color-shaded area under the characters acts as a visual label of the animation, while overlaps indicate blended motions. Best viewed in color.
    }
    \label{fig:qual}
\end{figure*}

We evaluate our blending method on two datasets, the Mixamo \cite{Mixamo} and the 100STYLE \cite{100Style}. For the Mixamo benchmark,  we select a subset of 24 joints from the original 65-joint skeleton to focus on the key motion patterns. For the 100STYLE dataset, since unedited motion captured BVH files are provided for each type of movement, we trim the sequences to retain only the segments that contain meaningful motion. Our model implementation is in PyTorch \cite{pytorch}, and all experiments are conducted on an NVIDIA GeForce RTX 4090 GPU. The training sequences range in length from 75 frames to 900 frames per motion, depending on the selected dataset, subsampled at 30 FPS. The training time is proportional to the training sequence length, \textit{e.g.,} taking about 3 hours to train our model on a motion sequence with around 360 frames and 15,000 iterations per level across all batches. The inference time is in the order of a few seconds, once the appropriate \textit{skeleton\_id\_maps} are given.

To evaluate the quality of the blended result, we follow the relevant literature \cite{rottshaham2019singan,li2022ganimator,SinMDM} and adopt the single-sample variant of the Fréchet Inception Distance (FID), the coverage (Cov), and the different diversity variants from \cite{li2022ganimator,SinMDM}: a) local (LDiv), b) global (GDiv), c) intra, and d) inter diversity. Additionally, we employ two L2-based metrics to assess the smoothness of the blended motion; the joint velocity and joint acceleration distances:

\begin{itemize}
    \item \textit{L2\_velocity} - Measures the difference in joint speed (\textit{i.e.,}~the L2 norm of the velocity vector) between two consecutive frames. Given the velocities $\mathbf{v}_{t,j} \in \mathbb{R}^3$ for each joint $j$ at each frame $t$, we first compute the L2 norm of the velocity vector for each joint and frame \(v_{t,j} = \left\| \mathbf{v}_{t,j} \right\|_2\). Then the \textit{L2\_velocity} is defined as: 
    \begin{equation}
        \Delta v_{t,j} = \left| v_{t,j} - v_{t-1,j} \right|.
        \label{eq:l2_velocity}
    \end{equation}
    
    \item \textit{L2\_acceleration} - To further analyze higher-order smoothness we measure the change in $\Delta v_{t,j}$ over time, which corresponds to temporal acceleration:
    \begin{equation}
        \Delta\Delta v_{t,j} = \left| \Delta v_{t,j} - \Delta v_{t-1,j} \right|.
        \label{eq:l2_acceleration}
    \end{equation}
\end{itemize}

For these metrics we consider information from five predefined joints: a) pelvis, b) left\_wrist, c) right\_wrist, d) left\_foot, and e) right\_foot. These joints are key indicators of potential discontinuities in the blended motion. By plotting these values for all frames, we can observe abnormalities at the blended region, indicating a non-smooth transition. In Fig.~\ref{fig:l2_metrics} we visualize the results using these two metrics for a blending motion on the 100STYLE dataset.

\subsection{Results}

\subsubsection{Quantitative analysis}

To study the performance of our model, we conduct a series of experiments focusing on the modulation mechanism and its integration across different levels of the generation process. The results of the ablation study on the Mixamo dataset \cite{Mixamo} are reported in Table~\ref{tab:quantitative_results}, while the ones on the 100STYLE dataset \cite{100Style} are presented in Table~\ref{tab:100_style_quantitative_results}. 

First, we investigate how the integration of the SPADE layer in generators of different levels affects performance. Specifically, we evaluate several configurations where the SPADE layer is applied on only one level of the hierarchy at a time. We experiment with placing the SPADE layer solely at the first, second, third, and fourth levels, respectively. Our results demonstrate that applying SPADE-like conditioning only at the first level yields the most favorable trade-off between low FID and high coverage in both datasets, resulting in consistent and smooth motion blending. This happens because the first level captures coarse motion features, which are critical for establishing the overall structure of the blended motion. 

Next, we experiment with an alternative normalization based modulation strategy. Specifically, we replace the SPADE module with a Feature-wise Linear Modulation (FiLM)~\cite{FiLM} layer. In this setup, instead of using skeleton-aware convolutional layers to produce the two tensors $\gamma$ and $\beta$, we compute them using skeleton-aware linear layers. However, as shown in the results, this configuration degraded performance by increasing the FID score both on the Mixamo dataset and 100STYLE, confirming the effectiveness of convolution-based spatial modulation. 

In Table~\ref{tab:quantitative_results} and Table~\ref{tab:100_style_quantitative_results}, we also present the results obtained by training GANimator \cite{li2022ganimator} on the same input motions, which does not include any conditional mechanism in the architecture. The results show the expected drop in the coverage metric for both datasets, as the generation process results in a random mix of the input motions.
Although our model exhibits lower diversity, this aspect is not our primary focus. In motion blending, diversity is secondary to generating smooth transitions between motions.

\subsubsection{Qualitative analysis}

Apart from the quantitative results, our qualitative evaluations further demonstrate the effectiveness of our model in generating smooth motion transitions. The visualizations in Fig.~\ref{fig:qual} highlight the blending quality achieved across different motion combinations from the Mixamo dataset~\cite{Mixamo}. 

The top example illustrates a sequence trained on the ``Capoeira'' and ``Breakdance Freezes'' sequences rendered using the ``Jackie'' character provided by Mixamo. Each input sequence contains approximately 250 frames, and we visualize subsampled unrolled frames in the top rows. In the final blended motion, we end up with a blended motion of 250 frames, as shown in the bottom row, where the first half of the frames are conditioned on the ``Capoeira'' input, while the second half correspond to the ``Breakdance Freezes'' motion. Notably, our model achieves smooth transitions even when the two input motions differ significantly in style and dynamics.

To further highlight the capabilities of our model, we present an additional example (bottom part of Fig.~\ref{fig:qual}) trained on three distinct motions: the ``Salsa\_Dancing'', the ``Swing\_Dancing'' and the ``Wave\_Hip\_Hop\_Dance''. For visualization purposes, we use the ``Michelle'' character provided by Mixamo. Each input sequence contains approximately 360 frames, and the full motion sequences are shown unrolled in the top rows. The bottom row displays the resulting blended motion. In this example, the final motion is conditioned on all three inputs. In this case, the first third of the sequence, \textit{i.e.,} 120 frames, follows the ``Salsa\_Dancing'' motion, while the second third reflects the ``Swing\_Dancing'' motion, and the final third continues with the ``Wave\_Hip\_Hop\_Dance''. As we can see, the model learns and combines distinct motion characteristics from each input motion, demonstrating that it can smoothly blend more than two motions.

\begin{figure}
    \centering
    \includegraphics[width=\columnwidth]{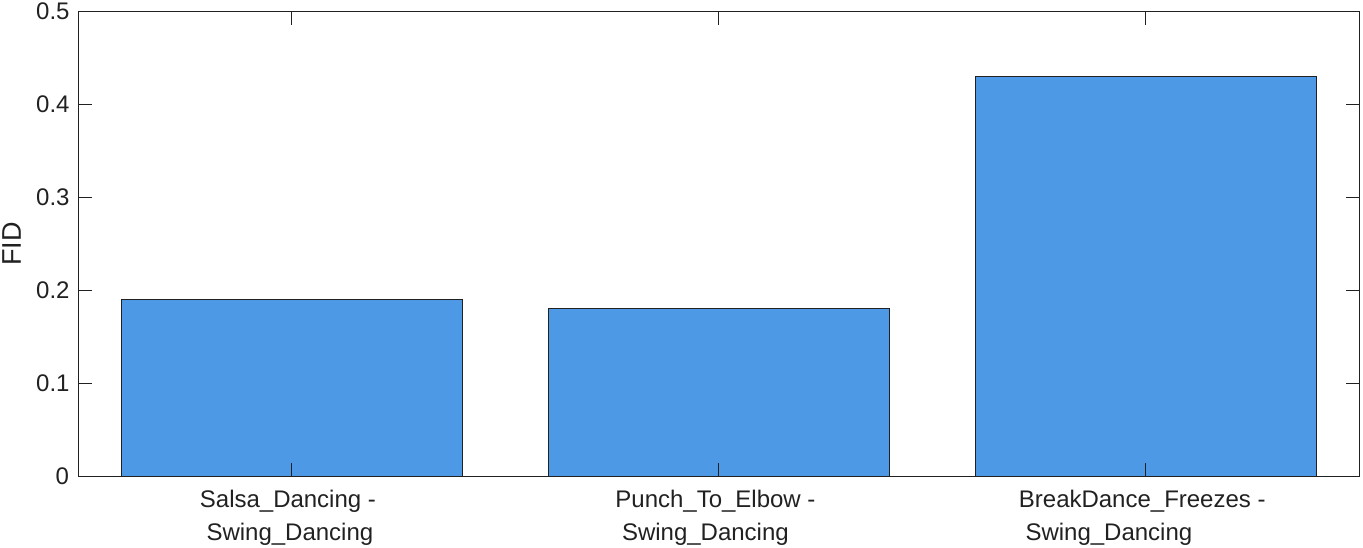}
    \caption{Blending quality as a function of motion similarity. We train our model on the ``Swing\_Dancing'' motion with three different motions: ``Salsa\_Dancing'', ``Punch\_To\_Elbow\_combo'', and ``Breakdance\_Freezes''. Results show that higher similarity between motions leads to lower FID scores.}
    \label{fig:exp}
\end{figure}

Finally, we also investigate how the similarity between different motions affects the quality of the generated blends. We hypothesize that the similarity between input motions plays a key role in determining the output quality. To explore this, we choose the motion sequence ``Swing\_Dancing'', from the Mixamo dataset and blend it with 3 different motion sequences: ``Salsa\_Dancing'', ``Punch\_To\_Elbow\_combo'' and ``Breakdance\_Freezes''. These motions are deliberately chosen to represent varying levels of similarity.
For instance, ``Breakdance\_Freezes'' differs significantly from ``Swing\_Dancing'', as it involves inverted body poses with feet off the ground, whereas ``Salsa\_Dancing'' and ``Punch\_To\_Elbow\_combo'' contain rhythmic and grounded movements, respectively, that align more naturally with the flow of ``Swing\_Dancing''. These shared motion patterns make it easier for the model to learn common features and blend the sequences effectively. For each configuration, we compute the FID score and present the results in Fig.~\ref{fig:exp}. As expected, the FID score is biased towards similarity when it comes to blending, and this is why the qualitative evaluation of generative animation models from experts (animators, 3D artists, etc.) is a necessity.

\section{Conclusion}
\label{sec:conclusion}
In this work, we introduced the first single-shot motion blending framework built upon a batched implementation of the GANimator model, augmented with a temporally-aware, SPADE-like conditioning mechanism. Our approach enables the controllable blending of two or more input animations within a single generative inference, producing results that are both visually plausible and kinematically smooth. This demonstrates the feasibility and effectiveness of learning to blend complex motions in a unified, efficient manner. Future work will investigate alternative conditioning strategies, such as textual or semantic prompts, to further enhance control and expressiveness in single-shot motion synthesis.
\newpage
{
    \small
    \bibliographystyle{ieeenat_fullname}
    \bibliography{main}
}

\end{document}